\documentclass[a4,twocolumn]{revtex4}

\usepackage{graphicx}
\usepackage{amsmath}
\usepackage{setspace}
\begin{document}

\title{RNA Folding Pathways in Stop Motion}
\author{%
Sandro Bottaro\,$^{1,*}$,
Alejandro Gil-Ley\,$^{1}$,
Giovanni Bussi \,$^{1,}$%
\footnote{To whom correspondence should be addressed.
Email: sbottaro@sissa.it, bussi@sissa.it
}}

\address{%
$^{1}$
Scuola Internazionale Superiore di Studi Avanzati, International School for Advanced Studies, 
265, Via Bonomea I-34136 Trieste, Italy
}

\begin{abstract}
 \noindent We introduce a method for predicting RNA folding pathways, with an application to the most important RNA tetraloops.  The method is based on the idea that ensembles of three-dimensional fragments extracted from high-resolution crystal structures are heterogeneous enough to describe metastable as well as intermediate states. 
  These ensembles are first validated by performing a quantitative comparison against available solution NMR data of a set of RNA tetranucleotides.
Notably, the agreement is better with respect to the one obtained by comparing NMR with extensive all-atom molecular dynamics simulations.
We then propose a procedure based on diffusion maps and Markov models that makes it possible to obtain reaction pathways and their relative probabilities from fragment ensembles.    This approach is applied to study the helix-to-loop folding pathway of all the tetraloops from the GNRA and UNCG families. The results give detailed insights into the folding mechanism
that are
compatible with available experimental data
and clarify the role of intermediate states observed in previous simulation studies.
The method is computationally inexpensive and can be used to study arbitrary conformational transitions.  
\end{abstract}

\maketitle

\section{Introduction}
Despite its simple 4-letters alphabet, RNA exhibits an amazing complexity, which is conferred by its ability to engage and interconvert between a variety of specific interactions with itself as well as with proteins and ions~\cite{tinoco1999rna}. A delicate balance between many different factors, such as hydrogen bonding, stacking interactions, electrostatics and backbone/sugar flexibility, determines structure and dynamics of RNA.  These interactions are explicitly described in atomistic molecular dynamics (MD) simulations, that represent an important computational tool for the investigation of RNA dynamics.  
Since the first MD simulation on tRNA~\cite{harvey1984phenylalanine}, the accuracy of atomistic force fields has steadily improved, to the point that
it is nowadays possible to obtain stable trajectories on the $\mu$s time scale for A-form double helices and tetraloops~\cite{banas2010performance,giambasu2015structural}. 
MD simulations have also proven useful to aid the interpretation of experimental data for structured RNAs~\cite{musiani2014molecular,pinamonti2015elastic} and protein-RNA complexes~\cite{whitford2013connecting,perez2015atp,krepl2015can}.
However, models used for nucleic acids are still significantly less accurate compared to those used for proteins~\cite{laing2011computational}. Recent simulations on systems amenable
to converged sampling showed that none of the current atomistic force fields correctly reproduces the behaviour of single stranded RNA tetranucleotides in solution~\cite{condon2015stacking,bergonzo2015highly}.
As a consequence, understanding the folding dynamics of small motifs such as RNA tetraloops
is extremely challenging using MD, both because of the force-field limitations and of the high computational cost.

From a different perspective, structural bioinformatics approaches to study RNA exploit the empirical relationship between sequence and three-dimensional structure.
These methods typically proceed by first extracting local conformations from a database of experimental structures, which are then assembled together to form complete models.  This idea lead to the development of successful fragment assembly algorithms for RNA ~\cite{Major1991,Gautheret1993,das2007automated,parisien2008mc} and proteins ~\cite{simons1997assembly}.
Fragment assembly methods often rely on the working hypothesis that frequencies of appearance in solved structures can be considered as an approximation to the underlying Boltzmann distribution. While this latter statement is in general not true, the statistics obtained from protein structures in the protein data bank (PDB) were shown to reproduce distributions as measured by NMR spectroscopy~\cite{best2006relation}
as well as quantum mechanical calculations~\cite{morozov2004close,butterfoss2003boltzmann}.
Additionally, the successful applications of statistical potentials to different RNA structure prediction problems suggest this approximation to be in practice acceptable~\cite{miao2015rna}.

By pushing these assumptions further, one could imagine the possibility to use fragment libraries to predict not only the most stable conformations but also intermediate and excited states~\cite{ee2014invisible}, so as to provide a description of reactive pathways.
For example, the conformational variations observed in crystal structures of RNA triloops have been linked to their internal dynamics~\cite{Lisi2007}, and a simple isomerization process occurring in the backbone of a small peptide was recently rationalised by analysing high-energy structures trapped in the PDB~\cite{brereton2015native}.
\enlargethispage{-65.1pt}

A further methodological step is however required to reconstruct the dynamics of systems undergoing large conformational changes,
possibly involving multiple pathways, when only an ensemble of structures at equilibrium is given.
Of particular interest in this context is the concept of diffusion map~\cite{coifman2005geometric}, which describes the long timescale dynamics
of a complex system using a transition matrix directly calculated from the data.
This important theoretical tool was applied to characterise the dynamics of proteins systems~\cite{rohrdanz2011determination}.
Interestingly, the slow eigenmodes of the transition matrix have been shown to be consistent with an analysis based on transition networks of long molecular dynamics simulations~\cite{zheng2011delineation}.  This result  suggests that the reactive pathways of macromolecular systems can be obtained from equilibrium ensembles alone.

In this Paper we show that the structural ensemble of RNA fragments extracted from the PDB exhibits remarkably good agreement with available NMR experimental data for five tetranucleotides. 
We then consider the fragment ensembles of the two most important families of RNA tetraloops: GNRA and UNCG (R=A/G and N=any). By using a formalism related to diffusion maps, we build a random walk on a graph in which each node is an experimental three-dimensional fragment and edges are weighted with an appropriate measure of similarity~\cite{bottaro2014role}.  
The properties of the resulting random walks are then analysed using the standard machinery of Markov state modeling~\cite{noe2009constructing} to obtain detailed folding trajectories.

We call the method for obtaining folding pathways from experimental fragments \emph{stop-motion modeling} (SMM), in analogy with the animation technique that produces the impression of motion through juxtaposition of static pictures.
SMM is a very general tool for obtaining transition pathways, it is computationally inexpensive, and it is therefore ideally suited to complement
MD simulations and to aid the interpretation of NMR spectroscopy data and kinetic experiments.

\section{Materials and Methods}
\textbf{Stop motion modeling} 
We here describe how to set up the stop motion modeling (SMM) procedure.\\
\textbf{1. Pairwise distances.} 
We first
calculate all pairwise distances between all 3D structures within an ensemble as
 $d_{ij} = \mathcal E$RMSD$(\mathbf{x}_i,\mathbf{x}_j) $.
Here $\mathbf{x}_i$ and $\mathbf{x}_j$ are the coordinates of structures $i$ and $j$ and $\mathcal{E}$RMSD is
an RNA-specific  metric based on the relative orientation of nucleobases only~\cite{bottaro2014role}.
This measure has the important property of being highly related to the temporal distance: this means that when two structures are close
in $\mathcal E$RMSD distance, then they are also kinetically close. 
In a previous work~\cite{bottaro2014role}, we have shown that this property is satisfied to a larger extent by $\mathcal E$RMSD compared to standard RMSD after optimal alignment~\cite{kabsch1976solution} as well as distance RMSD measures. Additionally, we have proven $\mathcal E$RMSD to be accurate in recognising known RNA motifs within the structural database.
$\mathcal E$RMSD  is also highly correlated with interaction network fidelity~\cite{Parisien2009}, and thus distinguishes structures with a different pattern of base-base interaction. \\
\textbf{2. Transition matrix.}
We build a graph on an ensemble of structures by constructing the adjacency matrix $K$ with Gaussian kernel $K_{ij} = \exp{(-d_{ij}^2/2\sigma^2)}$. Here, we set $\sigma=0.2$,
which is $\approx 1/3$ of the typical $\mathcal E$RMSD threshold used to consider two structures significantly similar.
This ensures that the transition probability among structures that exhibit a different pattern of base-base interaction is vanishingly small.
Following a procedure similar to the diffusion maps approach~\cite{coifman2005geometric}, and common in graph theory~\cite{chung1997spectral}, we construct the Markov matrix $T$ dividing by the diagonal degree matrix $D$, defined as $D_{ii} = \sum_j K_{ij}$.
To obtain a $T$ matrix that is simultaneously normalised and symmetric we introduce here an iterative normalisation procedure
$T_{(t+1)} = D_{(t)}^{-1/2}T_{(t)}D_{(t)}^{-1/2}$.  
Here the matrix product is implicit, the subscript $_{(t)}$ indicates the iteration, and $T_{(0)} = K$.
At convergence, this procedure yields a matrix $T$ that can be interpreted as a transition probability matrix.
Here $T_{ij}$ is the probability of observing a direct transition from state $i$ to state $j$.
$T$ is normalised ($\sum_j T_{ij}=1$)
and has an uniform equilibrium distribution over the ensemble of fragments ($\sum_i T_{ij}=1$).
This procedure is closely related to the most common version of graph Laplacian normalisation.
However, we notice that in the usual procedure the transition matrix is made non symmetric by
normalisation, and thus has an equilibrium distribution that is not necessarily uniform.
The advantage of our formulation is that the
averages computed from the random walk are by construction identical the to ensemble averages obtained from
the original set of structures.
\\
\textbf{3. Transition pathways between states.}
Dynamical properties of the system are calculated from the transition matrix $T$ as described in Ref.~\cite{noe2009constructing}, and briefly reported in Supporting Information 1 (SI1) for clarity. 
The flux calculations are performed using pyEMMA \cite{scherer_pyemma_2015}.
To facilitate the analysis of the fluxes, nodes are  lumped together using a standard spectral clustering procedure~\cite{ng2002spectral} on the transition matrix $T$.
Notice that the lumping is only used to compute the aggregate fluxes, and does not influence in any way the underlying Markov model.
We set the number of clusters to 25 and 45 for UNCG and GNRA tetraloops, respectively.  
We empirically verified that the fluxes are robust with respect to the number of clusters.
The flux between distant structures (compared to the Gaussian width $\sigma$) depends on the number of transition pathways connecting them.
If the overall transition requires crossing intermediate states with very low population, corresponding to high free-energy barriers, the resulting flux will be very small.
In the limit of missing intermediates, the graph becomes disconnected and the resulting flux is zero.
\\
\textbf{4. Low-dimensional embedding.} For the sake of visualisation, folding pathways and clusters are projected on a low dimensional
space. To this aim we use the diffusion map technique, where the top eigenvectors of the matrix $T$ are interpreted
as coordinates that provide a low-dimensional embedding in which the local structure (small distances)
is preserved~\cite{coifman2005geometric,rohrdanz2011determination}. We empirically found that to make
the visualisation clearer it is convenient to use a value of  $\sigma$ 3-4 times larger compared to the one used to calculate the fluxes. This choice only affects the two-dimensional projection, and not the calculated pathways and clusters.

\textbf{Molecular dynamics simulations.} Molecular dynamics simulations were performed using the GROMACS software package~\cite{pronk2013gromacs}. RNA was modeled using the Amber99 force field with parmbsc0 and $\chi_{\text{OL3}}$ corrections and was solvated in explicit water and
ions~\cite{cornell1995second,perez2007refinement,banas2010performance,jorgensen1983comparison,joung2008determination}. Parameters are available at http://github.com/srnas/ff. Temperature replica exchange MD was used to accelerate sampling~\cite{sugita1999replica}. For each system 24 replicas in the temperature range 300K-400K were simulated for 2.2 $\mu$s per replica. More details are available in SI2. \\

\section{Results}

We extract all the fragments with a given 4-nucleotide sequence from high-resolution crystal structures in the PDB. 
We consider all RNA-containing structures with resolution $3.5$\AA\ or better available in the PDB database as of 8/19/2015, for a total of 1882  structures.
A complete list is provided in SI3.
This procedure yields a collection of structures (fragments ensemble) composed by 2-15 thousands fragments, depending on the sequence. 

\subsection{Comparison against NMR data}
Recent NMR spectroscopy studies on RNA tetranucleotides showed  AAAA, CCCC, CAAU and GACC to be mostly in A-form in solution, while data for UUUU were compatible with  more disordered, partially stacked conformations~\cite{condon2015stacking,tubbs2013nuclear}. In all cases, no evidence of intramolecular hydrogen bonding was found.

\begin{figure}
\centerline{\includegraphics[scale=1.0]{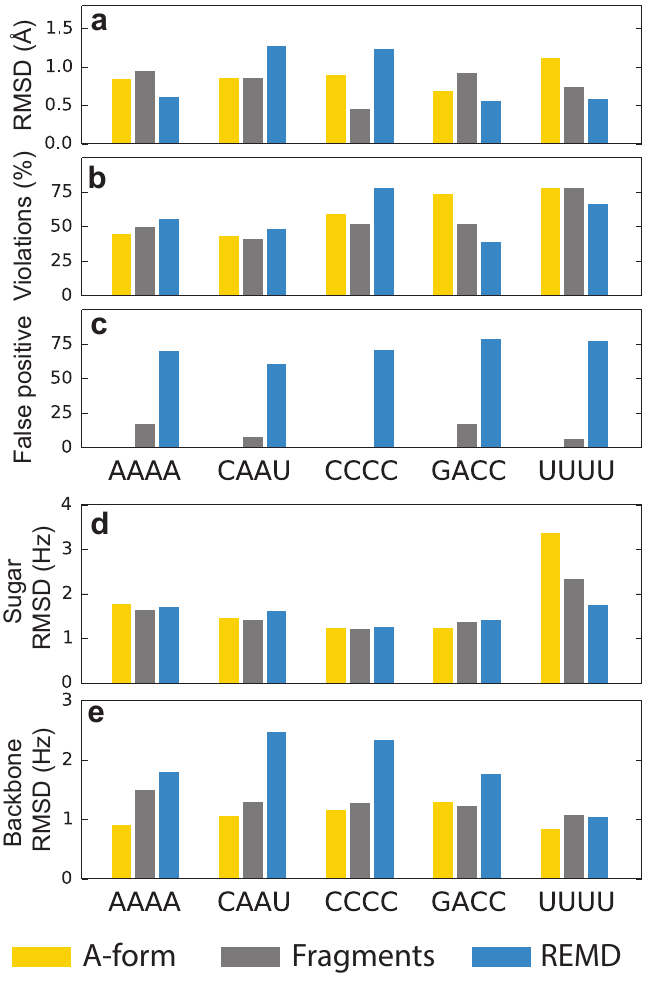}}
\caption{Comparison between calculated and experimental NMR data. Calculations were performed on a single ideal A-form helix (yellow bars), on the ensemble of fragments extracted from the PDB (grey bars), and on replica exchange molecular dynamics simulations (REMD, blue bars). \textbf{a}) RMSD between calculated and predicted average NOE distances. \textbf{b}) Percentage of predicted NOE average distance outside the experimental range.   \textbf{c})  Number of false positives, \textit{i.e.} predicted distances below 5 \AA\ not observed in experimental data.  \textbf{d}) RMSD between experimental and predicted $^3\text{J}_{1'2'},^3\text{J}_{2'3'},^3\text{J}_{3'4'}$ scalar couplings, reporting on sugar pucker geometry. \textbf{e})  RMSD between experimental and predicted $^3\text{J}_{5'/5''P},^3\text{J}_{4'5'/5''},^3\text{J}_{3'P}$ scalar couplings, reporting on backbone geometry. \label{noes}}

\end{figure}

In this section we compare available NMR data with ideal A-form helices, with ensembles of fragments from the PDB, and with MD simulations.  As a term of reference, we first set out to compare available experimental NMR data with the prediction obtained from ideal A-form helices.
Figure \ref{noes} shows the agreement of nuclear Overhauser effect (NOE) data with the values predicted from A-form helices in terms  of a)\ root mean square deviation (RMSD), b) percentage of NOE distance violations and c) number of non-local false positives, \textit{i.e.} distances between protons in non-consecutive nucleotides predicted to have average NOE distance $\le 5$ \AA\ but not visible in the experiment. These definitions are consistent with those used in Ref.~\cite{condon2015stacking}.
The deviation between predicted and experimental data is in general low (Fig. \ref{noes}a).
However, since flexible tetraloops can adopt non-A-form structures, a large fraction of predicted distances falls outside the experimental range (Fig. \ref{noes}b).
Furthermore, systematic discrepancies between NMR spectroscopy and X-ray crystallography have been discussed~\cite{gendron2001quantitative} and contribute to the fraction of NOE violations. In all cases, no false positive is observed.  Compatibly with its more disordered behaviour in solution, the predicted NOE distances for the A-form UUUU is poorer compared to the other tetranucleotides. 
In Fig.\ref{noes} we show the agreement between experimental and predicted $^3$J scalar couplings, considering data reporting on sugar pucker (panel d) and on  backbone conformation (panel e). As observed for NOE, experimental scalar couplings are compatible with A-form helices, with the exception of UUUU. 
In the latter case, the experimental data suggest high sugar mobility, with significant deviation from the common C3'-endo sugar pucker conformation.

The accord of the fragment ensembles with experimental data is very similar to what observed for the A-form helix (Fig.\ \ref{noes}).
This can be rationalised at least for the first four tetranucleotides by considering that Watson-Crick double helices are the dominating conformations in the crystal structures deposited in the PDB database. In the case of the UUUU tetranucleotide, the conformational ensemble obtained from the fragments improves
the agreement with NMR data when compared with the ideal A-form helix. This highlights the importance of using a diverse ensemble of structures to describe a flexible molecule in solution.

Figure \ref{noes} also reports the agreement between experimental data and temperature replica exchange molecular dynamics (REMD)~\cite{sugita1999replica} at atomistic resolution.
In terms of RMSD, NOE distances predicted by REMD simulations are comparable with those obtained from the fragments ensemble.
However, a high number of non-local NOE false positives can be observed for all the systems (Fig.\  \ref{noes}c) .
As reported in recent MD studies, AMBER force fields over-stabilise intercalated structures with stacking between bases 1-3 and 1-4 ~\cite{condon2015stacking,bergonzo2015highly}.
These structures, which are also observed in our simulations, cause the appearance of spurious contacts  that are not compatible with experimental NOE distances (\textit{i.e.} false positives). 
The wrong pattern of stacking interactions observed in REMD simulations affects the backbone conformation as well, resulting in a poor agreement with  backbone $^3$J scalar coupling data (Fig.\  \ref{noes}e). 
With respect to sugar pucker, instead, it is worth noting that for UUUU both C3'-endo and C2'-endo conformations are significantly populated in REMD simulations, in accord with scalar coupling data (Fig.\ \ref{noes}d).  Scatter plots of experimental versus predicted NOEs and $^3$J scalar couplings are shown in SI4 and SI5.

Taken together, these results show that the fragments ensembles are overall in accord with NMR data. For tetranucleotides which in solution adopt the A-helix form,
the agreement is comparable with what obtained from a single structure.
On the contrary, noticeable artifacts are observed in MD simulations, in agreement with previously reported results where all the state-of-the-art
force fields were tested~\cite{condon2015stacking,bergonzo2015highly}.
The most striking discrepancy between simulations reported here and NMR is caused by the presence of intercalated structures. %
This indicates that results obtained sampling the fragment ensemble could be potentially more accurate than expensive MD simulations in reproducing
biomolecular dynamics, at a fraction of the computational cost.

\subsection{Stop-motion modeling: mimicking dynamics using static snapshots}
The results described above show that the conformational ensemble of fragments on 5 different systems is in agreement with NMR data.
Building upon this result, we use the fragments ensemble to study the dynamical properties of arbitrary RNA sequences. 
First, we assume that the fragments ensemble for a given sequence represents the equilibrium distribution at some temperature. 
We then construct a Markov matrix  on all the fragments within the ensemble. 
Following a procedure closely related to the one used in the construction of diffusion maps~\cite{coifman2005geometric}, we calculate transition probabilities as Gaussian function of their distance.
In the present context, distances are measured using the $\mathcal{E}$RMSD~\cite{bottaro2014role}, which is based on the relative position and orientation of nucleobases only. 
The resulting Markov matrix contains kinetic information, from which it is possible to either generate reactive trajectories with a stochastic procedure or
to analyse the associated fluxes as it is customarily done for Markov state models~\cite{dellago1998efficient,noe2009constructing}.
 We call this procedure stop-motion modeling (SMM).
 In the next section we present the results of the SMM on different tetraloops, while we refer the reader to the Materials and Methods section for an in-depth, technical description of the algorithm. 

\subsection{Folding pathways of RNA tetraloops in stop motion}
We use the SMM approach described above to study the helix-to-loop transition of UNCG and GNRA tetraloop families (see Table 1). When stabilised by additional Watson-Crick base pairs, these sequences are known to adopt specific stem-loop structures~\cite{hall2015mighty}.  The corresponding fragments ensembles are indeed typically composed by i) fully stacked structures in A-form conformations, ii)  folded tetraloops and iii)  other conformations that are distant from both loop and A-form.
\begin{table}
\caption{Number of structures within each fragments ensemble. The fraction of loops and A-form helices is calculated considering all structures with $\mathcal{E}$RMSD$<$0.6 from the consensus loop/ ideal helix. }
\begin{footnotesize}
\begin{tabular}{@{\vrule height 9pt depth4pt  width0pt}lccc}
  Sequence & \# of structures & Loop (\%) & A-form (\%)\\
  \hline
UACG & 3674 & 9 & 37 \\
 UCCG & 6673 & 3 & 67 \\
UGCG &  5872  & 0 & 61 \\
UUCG  & 3969  & 15 & 17 \\
\hline
GAAA & 12908 & 16 & 8 \\
GAGA & 7596 & 10 & 27 \\
GCAA &  7835 & 14 & 23 \\
GCGA & 10946 & 7 & 17 \\
GGAA &  12395 & 3 & 24 \\
GGGA & 16764 & 0.7 & 46 \\
GUAA &  8100  & 6 & 15 \\
GUGA & 11432 & 10 & 10 \\
\end{tabular}
\end{footnotesize}
\end{table}

Here we assume that the folding mechanism is determined by the tetraloop only, and we do not model the full stem-loop sequence.
As a consequence, our investigation only provides insight on the folding mechanism of the analysed tetraloop sequences,
without considering possible effects of the stem sequence.

\begin{figure*}
\centerline{\includegraphics[scale=1.0]{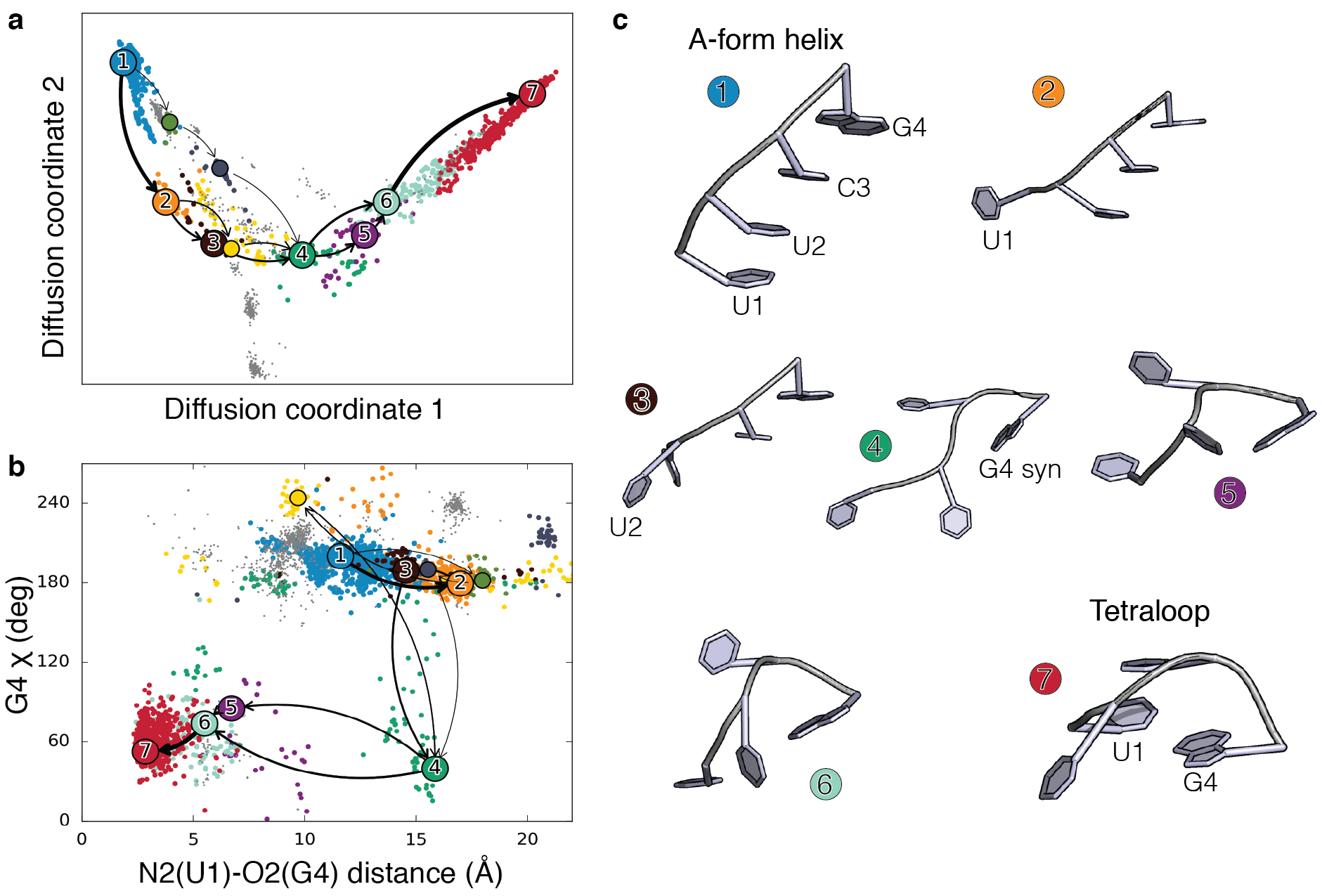}}
\caption{\textbf{a}) UUCG Helix to loop folding pathways projected on the first two diffusion coordinates. Clusters corresponding to A-form helix and to the loop  are coloured in blue and red, respectively.  On-pathways are shown in colours, while grey points correspond to clusters not contributing to the folding pathway. Arrows show the dominant folding pathways ($\approx$ 75\% of the total folding flux). Line width is proportional to the flux. \textbf{b}) Helix to loop  folding pathway projected on the distance between atom O2 in base U1 and atom N1 in base G4 vs the glycosidic torsion angle in G4.
The distance reports on the formation of a signature interaction of the tetraloop.
\textbf{c}) Centers of the clusters in the main folding pathway, numbered and coloured as in panels a-b. \label{UUCG}}
\end{figure*}

\textbf{UNCG Tetraloop.} The UNCG is one of the most abundant and well-characterised families of tetraloops~\cite{tuerk1988cuucgg}.
NMR and X-ray structures of these tetraloops revealed that their high stability is conferred by a peculiar
trans-Watson-Crick/sugar-edge (tWS) base pair~\cite{leontis2001geometric} between G4 and U1, together with extensive U2-G4 stacking and a U2-C3 base-phosphate interactions~\cite{cheong1990solution,ennifar2000crystal}. 
In Fig.\ \ref{UUCG} we show the main folding pathways obtained from SMM connecting the A-form helix (cluster 1) to the UNCG tetraloop (cluster 7).
For visualisation purpose, the folding pathways are projected on the first two diffusion coordinates, that describe the slowest directions
of propagation of the Markov chain~\cite{coifman2005geometric}.

Additionally, we project the folding pathways on the U1-G4 distance and on the value of  the $\chi$ torsion angle in G4, reporting respectively on U1/G4 base-pair formation and on the \textit{anti/syn} transition of the G4 glycosidic bond (Fig.\ \ref{UUCG}b). 
For UUCG we found three dominant folding pathways. In the first one (36\% of the total flux) we observe an initial elongation phase, during which U1/U2 unstack (cluster 2), followed by the loss of U2/C3 stacking interaction (cluster 3).
The loop then undergoes a major rearrangement, in which G4 flips into \textit{syn} conformation (cluster 4), U1 and G4 approach (clusters 5-6), until the formation of the tetraloop, featuring the characteristic G4-U1 tWS base pair (cluster 7). 
The second pathway (23\% of the flux) is qualitatively similar to the first one, as most of the intermediates are in common. 
In the third pathway (17\% of the flux), instead, unstacking first occurs at the 3' end.  A stochastic trajectory representing the folding process is shown in Supplementary Movie 1.

UACG tetraloop folding proceeds through a similar mechanism, although G4 flipping can also occur after loop compaction, as shown in
 SI6. 
 The SMM analysis could not be performed on the UCCG and UGCG sequences because none or very few of the fragments analysed adopt the tetraloop conformation (Table 1).

It is worthwhile observing that the remaining $\approx 25\%$ of the reactive flux visits not only the clusters discussed above, but also other fragments (shown as grey dots in Fig.\  \ref{UUCG}).
It is however possible to calculate the contribution of pathways visiting specific conformations of interest.
In a previous MD simulation study it was identified a metastable UUCG tetraloop in which G4 is in \textit{anti} conformation~\cite{kuhrova2013computer}.
This structure was discussed in terms of its similarity to loop 32-37 in PDB 3AM1~\cite{sherrer2011c}. In our analysis, this specific structure from the PDB is assigned to one of the
least populated clusters (SI7). The pathways visiting this cluster only contribute to a
very small fraction of the reactive flux ($\approx 0.2\%$), suggesting that this could be an off-pathway intermediate.

\begin{figure}
\centerline{\includegraphics[scale=1.0]{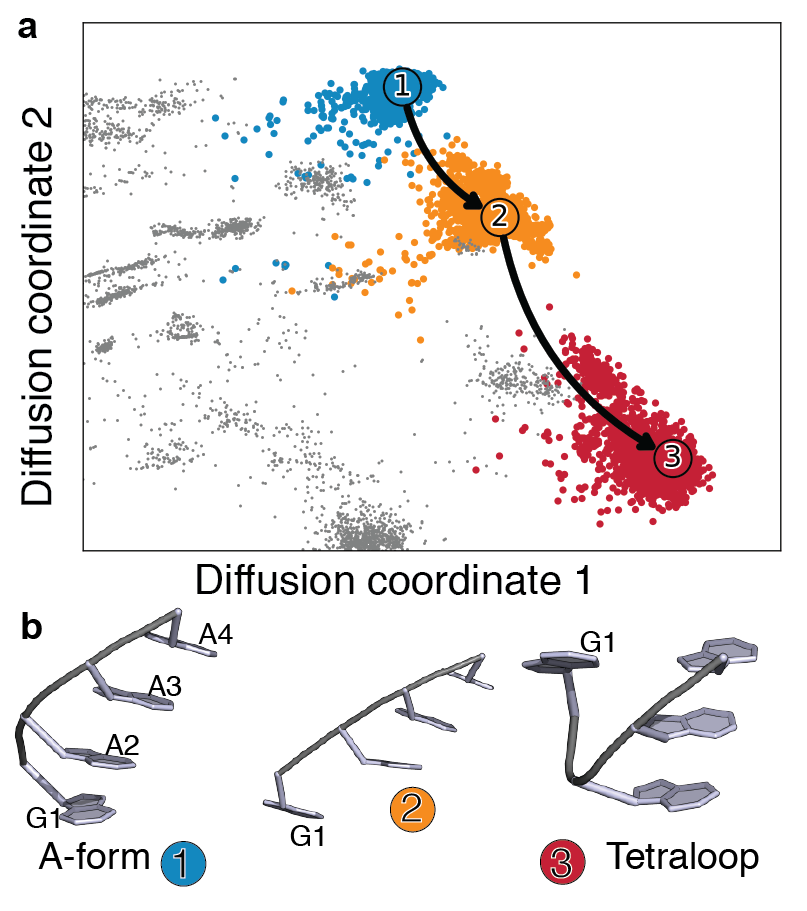}}
\caption{\textbf{a}) GAAA A-form helix to loop folding pathways projected on the first two diffusion coordinates. The first folding pathway, contributing for more than 90\% of the total flux, is indicated as a solid line. Only the part of the diffusion map containing the clusters that contribute significantly to the folding pathways is shown. See SI8 for the full two-dimensional projection.  \textbf{b}) Centers of the clusters in the main folding pathway, numbered and coloured as in panel a.  \label{GAAA}}
\end{figure}

\textbf{GNRA Tetraloop}
Contrary to UNCG, GNRA tetraloops often mediate RNA tertiary interactions~\cite{michel1990modelling}. GNRA tetraloops are characterised by a trans sugar/Hoogsteen (tSH) G1-A4 non-canonical base-pair, and by N2-R3-A4 stacking~\cite{heus1991structural}. We show in Fig.\ \ref{GAAA} the folding pathways obtained from SMM on the GAAA tetraloop. Starting from a fully stacked A-form helix (cluster 1), the main folding pathway consists in two steps, namely  G1 unstacking at 5' end (cluster 2) followed by a rotation around the $\alpha$ dihedral angle in A2~\cite{westhof1989computer}, leading to the formation of the tSH G4-A1 base pair (cluster 3). For GAAA, as well as for GAGA tetraloops (see SI8 and SI9) this pathway contributes for $90\%$ of the flux.

\begin{figure}
\centerline{\includegraphics[scale=1.0]{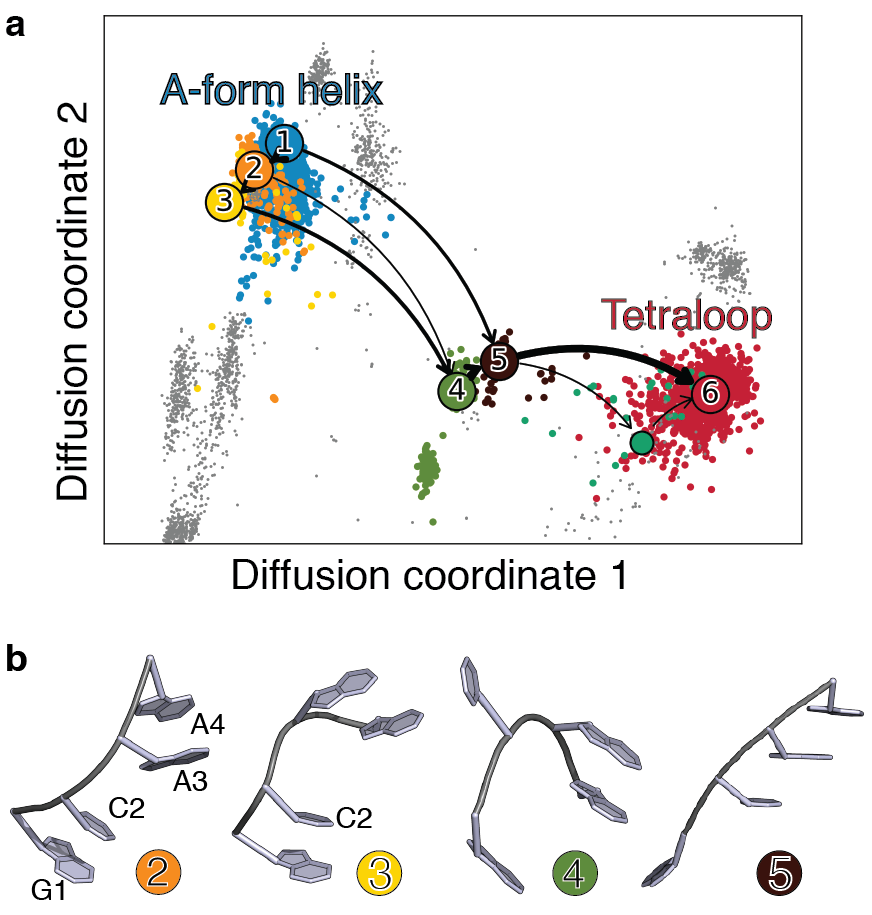}}
\caption{\textbf{a}) GCAA helix to loop folding pathways projected on the first two diffusion coordinates. 
 Clusters corresponding to A-form helix and to the loop  are coloured in blue and red, respectively.  On-pathways are shown in colours, grey points correspond to fragments not contributing to the folding pathway. The first three folding pathways, contributing  for $\approx 80\%$ of the total flux, are indicated with black arrows. Line width is proportional to the flux.   \textbf{b}) Centers of the clusters in the main folding pathway, numbered and coloured as indicated in panel a. The three-dimensional structures of the A-form helix  (cluster 1) and tetraloop (cluster 6) are not shown, as they are equivalent to clusters 1 and 3 in Fig. \ref{GAAA}. \label{GCAA}}

\end{figure}

When considering the third most common GNRA tetraloop sequence (GCAA)  a folding mechanism similar to the one described above is observed, with the additional presence of stacking/unstacking dynamics between C2 and A3, as shown in Fig.\ref{GCAA}, clusters 1-3.   More generally, we systematically observe significant unstacking dynamics in all GYRA tetraloops (Y=C or U) which is absent in GAAA and GAGA tetraloops (see SI9). This is expected, as purine-pyrimidine stacking is less strong compared to purine-purine stacking.  
Note that the high resolution structures used here do not contain a significant percentage of GGGA and GGAA sequences forming the consensus GNRA tetraloop (Table 1), thus making it difficult to perform the SMM on these two sequences. 

\textbf{Diffusion maps identify kinetically distant motifs}
The Euclidean distance on the diffusion map is related to the rate of connectivity of the points in the Markov chain~\cite{coifman2005geometric}. 
Therefore, the diffusion map can be used to identify structures which are kinetically far from each other, irrespectively of their relevance in the helix-to-loop transition.
As an example, we report the presence of structures in the GAAA fragment ensemble featuring the typical signature interactions of the UUCG tetraloop (see SI8).
This similarity has not been reported before and 
suggests that some tetraloops with GNRA sequence could have a functional role similar to the one of UNCG tetraloops.

\section{Discussion} 

In this Paper we present a method to build RNA folding pathways based on the analysis of the ensemble of  fragments in the available structural database.
The different conformations and their frequencies are dictated by many factors: the available experimental structures, the crystallisation conditions, the crystal packing, and the non-local interactions with proteins, ions, and other RNA bases. All these differences act as perturbations that stabilise intermediate states and alternate minima.

Firstly, we show that the fragments ensembles quantitatively agree with available NMR spectroscopy data for 5 selected RNA tetranucleotides.  This result is compatible with a previous study showing that the structural ensembles of  proteins in the PDB  provide a representative sampling of the heterogeneity of their native states as probed by various NMR measurements~\cite{best2006relation}.
The agreement is a non trivial result since the PDB has an intrinsic database bias, in which A-form helices are likely over-represented with respect to other structures.
It is important to mention that this bias could be alleviated or removed by using filtering procedures based on available experimental data ~\cite{salmon2013general,emani2014elucidating}.
In principle, one could even extrapolate the structural ensembles to different temperatures or ionic conditions, provided that enough overlap between distributions exists.
We also performed extensive, fully atomistic MD simulations showing instead artifacts not compatible with NMR data for the same tetranucleotides.
This confirms the results obtained in recent simulation studies~\cite{condon2015stacking,bergonzo2015highly}.

Secondly, we extend the scope of fragments ensemble from thermodynamics to dynamics. More precisely, 
we introduce a procedure, called stop-motion modeling (SMM), for analysing structural ensembles, so as to produce reaction pathways for generic conformational transitions in RNA.
This procedure finds its theoretical underpinnings in the framework of transition-path theory~\cite{dellago1998efficient} and Markov-state models~\cite{noe2009constructing}, widely used in the protein simulation community~\cite{chodera2007automatic}. Here, however, we introduce two significant modifications.  First, we use experimental crystal fragments instead of molecular dynamics to construct the Markov states. Second, we calculate transition probabilities from structural distances instead of estimating rates from many, short MD simulations. This latest idea is borrowed from the framework of diffusion maps, where a transition matrix is obtained from structural distances~\cite{rohrdanz2011determination}.

 We employ SMM to study the helix-to-loop transition of UNCG and GNRA tetraloop families, leading to a detailed description of their folding pathways.
This study would be very difficult or not possible using atomistic MD simulations
as well as other structural bioinformatics approaches~\cite{das2007automated,frellsen2009probabilistic}, due to the known issues
in properly modeling RNA tetraloops.
 Note that the stability of the full tetraloop structures is known to be dependent on the length and sequence of the stem. Here we combined fragments with a fixed
 sequence in the loop without explicitly considering the nucleotides of the stem, as our procedure relies on the fact that the heterogeneity of sequences
 observed in the PDB  acts as a perturbation apt to stabilise the most accessible intermediate states.
 As an example, by exclusively considering fragments with sequence cUUCGg, the ensemble would consist almost entirely of
 folded tetraloops. The inclusion in the ensemble of fragments with sequence cUUCGc has the net effect of destabilising the loop,
 making it possible to observe extended conformations. 

 For the UUCG tetraloop we observe a folding mechanism characterised by a progressive unstacking followed by 
 a concerted movement involving \textit{anti}  to \textit{syn} flip of the glycosidic bond and loop compaction.
 T-jump experiments  suggested a four-states sequential folding model characterised by
unstacked (S), unfolded (U), frayed (E), and native (N) state~\cite{ma2006exploring,sarkar2009fast}.
We hypothesise the unfolded (yet stacked) state U to correspond to clusters 2-3 in Fig. \ref{UUCG}, and 
the frayed state to correspond to the compact clusters 5-6.
In T-jump experiments the  unfolded and unstacked state S is populated only at high temperatures. Consistently, this state is not observed
in the fragment ensemble. Our analysis also shows the G4 \textit{anti}$\to$\textit{syn} flip to occur concertedly with stem formation, 
with no direct evidence suggesting the tetraloop folding to occur from a misfolded, compact structure.
It is important to observe that structures where the stem is formed and G4 is in \textit{anti} were observed in previous MD simulation studies~\cite{kuhrova2013computer,chen2013high}.
Although similar structures were also present in the PDB and thus included in our fragment ensembles, the pathways visiting these conformers
contribute to a small extent to the reactive flux, suggesting  these structures to be off-pathway intermediates.

Tetraloops of the GNRA family show a simpler folding mechanism, in which G1 unstacks from N2 and then forms a base pair with A4. Fluorescence decay experiments~\cite{zhao2007direct} as well as NMR measurements~\cite{jucker1996network} strongly suggest that GNRA tetraloops undergo significant conformational dynamics, in which N2 and R3 can interconvert between different stacking arrangements. In particular, C2 in GCAA tetraloop was found to be much more dynamic with respect to A2 in GAGA and GAAA. This is completely consistent with our analysis that shows significant stacking dynamics in GCAA (Fig.\ \ref{GCAA}), and more generally in sequences where N2 is a pyrimidine. 

The results presented in this Paper show fragment ensembles to provide a quick and realistic manner to generate equilibrium distributions for short RNA sequences compatible with
solution experiments.
The introduced  SMM procedure makes it possible to obtain reaction pathways from these ensembles at a small computational cost.
This allows us to provide a detailed description of the folding mechanism for the most common RNA tetraloops 
that is compatible with available experimental data and clarifies the role of intermediate states observed in previous MD studies.
In this work we focused on systems for which the statistics of fragments extracted from the PDB
is sufficient to generate paths between relevant metastable conformations. The SMM procedure can be straightforwardly applied
to larger systems provided that meaningful ensembles are available.
We envision the possibility of using SMM to analyse ensembles generated by MD or fragment-assembly techniques~\cite{das2007automated,parisien2008mc} for the characterisation of conformational dynamics in larger molecules.

\section{Availability}
Stop motion modeling has been implemented in the baRNAba package freely available at https://github.com/srnas/barnaba.

\section{Acknowledgments}
Pascal Auffinger, Pavel Banas, Alan Chen, Richard Andre Cunha, Kathleen Hall and Eric Westhof are acknowledged for reading the manuscript and providing useful comments.

\section{Funding}
The research leading to these results has received funding from the European Research Council under the European Union's Seventh Framework Programme (FP/2007-2013) / ERC Grant Agreement n. 306662, S-RNA-S.

\end{document}